\documentclass[prd,aps,twocolumn,a4paper,floatfix,nofootinbib]{revtex4}

\usepackage[utf8]{inputenc}
\usepackage{graphicx,psfrag}
\usepackage{mathrsfs}
\usepackage{amsmath,amsfonts,amssymb}
\usepackage{multirow}
\usepackage{diagbox}
\usepackage{comment}
\usepackage{xcolor}
\usepackage{enumerate}
\usepackage{booktabs}
\usepackage[normalem]{ulem}

\usepackage{hyperref}
\hypersetup{
    colorlinks = true,
    linkcolor = {blue},
    citecolor = {blue},
    urlcolor = {blue},
    linkbordercolor = {white},
    }

\usepackage{color}
\definecolor{cyan}{rgb}{0,0.9,0.9}
\definecolor{orange}{rgb}{0.9,0.5,0}
\definecolor{magenta}{rgb}{1,0,1}
\definecolor{purple}{rgb}{0.8,0.4,0.8}
\definecolor{gray}{rgb}{0.8242,0.8242,0.8242}
\definecolor{green}{rgb}{0.,0.8,0.}

\usepackage[normalem]{ulem}

\begin{document}

\title{Long-Term Simulations of Dynamical Ejecta: Homologous Expansion and Kilonova Properties}

\author{Anna \surname{Neuweiler}$^{1}$}
\author{Tim \surname{Dietrich}$^{1,2}$}
\author{Mattia \surname{Bulla}$^{3,4}$}
\author{Swami Vivekanandji \surname{Chaurasia}$^{3}$}
\author{Stephan \surname{Rosswog}$^{3,5}$}
\author{Maximiliano \surname{Ujevic}$^{6}$}

\affiliation{${}^1$Institut f\"{u}r Physik und Astronomie, Universit\"{a}t Potsdam, Haus 28, Karl-Liebknecht-Str. 24/25, 14476, Potsdam, Germany}
\affiliation{${}^2$Max Planck Institute for Gravitational Physics (Albert Einstein Institute), Am M\"uhlenberg 1, Potsdam 14476, Germany}
\affiliation{${}^3$The Oskar Klein Centre, Department of Astronomy, Stockholm University, AlbaNova, SE-10691 Stockholm, Sweden}
\affiliation{${}^4$Department of Physics and Earth Science, University of Ferrara, via Saragat 1, 44122 Ferrara, Italy}
\affiliation{${}^5$Sternwarte Hamburg, Gojenbergsweg 112, 21029 Hamburg, Germany}
\affiliation{${}^6$Centro de Ciências Naturais e Humanas, Universidade Federal do ABC, Santo André 09210-170, SP, Brazil}

\date{\today}

\begin{abstract}
Accurate numerical-relativity simulations are essential to study the rich phenomenology of binary neutron star systems. In this work, we focus on the material that is dynamically ejected during the merger process and on the kilonova transient it produces. Typically, radiative transfer simulations of kilonova light curves from ejecta make the assumption of homologous expansion, but this condition might not always be met at the end of usually very short numerical-relativity simulations. In this article, we adjust the infrastructure of the \textsc{bam} code to enable longer simulations of the dynamical ejecta with the aim of investigating when the condition of homologous expansion is satisfied. In fact, we observe that the deviations from a perfect homologous expansion are about $ \lesssim 30$\% at roughly $100$\,ms after the merger. To determine how these deviations might affect the calculation of kilonova light curves, we extract the ejecta data for different reference times and use them as input for radiative transfer simulations. Our results show that the light curves for extraction times later than $80$\,ms after the merger deviate by $\lesssim 0.4$\,mag and are mostly consistent with numerical noise. Accordingly, deviations from the homologous expansion for the dynamical ejecta component are negligible for the purpose of kilonova modelling.
\end{abstract}

\maketitle

\section{Introduction}
\label{sec:intro}

About 90 gravitational wave (GW) events have been detected since the Advanced LIGO and Advanced Virgo detectors started operating~\citep{LIGOScientific:2018mvr,LIGOScientific:2020ibl,LIGOScientific:2021}. The observed GWs originated from the merger of three different systems: binary black holes (BBHs), e.g.,~\citep{Abbott:2016blz,Abbott:2016nmj}, binary neutron stars (BNSs)~\citep{TheLIGOScientific:2017qsa, Abbott:2020uma}, and black hole - neutron stars (BHNSs)~\cite{LIGOScientific:2021qlt}. Systems containing at least one neutron star (NS) are especially interesting, as additional electromagnetic (EM) counterparts might be present. The observation of a phenomenon via different messengers can provide valuable insights into both the involved physics and the astronomical environment. The first and so far only event for which GW and EM counterparts were unambiguously detected was GW170817 which occurred on August 17$^\mathrm{th}$, 2017~\citep{TheLIGOScientific:2017qsa}. The kilonova AT2017gfo~\citep{Abbott:2017wuw} and the gamma-ray burst GRB170817A~\citep{Goldstein:2017mmi, Savchenko:2017ffs} are associated with the same source. \par

Because of their high compactness, NSs allow for the study of matter at densities that are inaccessible to terrestrial laboratories. In the last few decades, NS mergers have been considered the place for the formation of the heaviest elements in our Universe~\cite{Lattimer:1974a, Eichler:1989ve, Freiburghaus:1999, Rosswog:1998hy, Korobkin:2012uy, Wanajo:2014wha}. The formation of about half of all elements heavier than iron involves neutron capture reactions that must be rapid (r-process) compared to $\beta$-decay~\cite{Burbidge:1957vc, Cameron:1957}, and therefore requires a neutron-rich environment, see \cite{cowan21} for a recent review. Analogous to type Ia supernovae, the radioactive decay of the formed r-process nuclei is expected to power an EM transient in the optical, infrared, and ultraviolet bands. In the literature this transient is called a kilonova~\citep{Li:1998bw, Metzger:2016pju} or macronova~\citep{Kulkarni:2005jw}. Indeed, the EM counterpart AT2017gfo showed signatures indicative of a transient triggered by r-process nuclei including Lanthanides and Actinides~\cite{Barnes:2016umi, Kasen:2017sxr, Tanaka:2017qxj, Tanvir:2017pws, Rosswog:2017sdn, wu19, Miller:2019dpt, kasliwal22}. \par 

Simulations based on numerical-relativity (NR) are crucial for the study of these systems and a correct interpretation of the observational data. By solving Einstein's field equations, NR enables accurate simulations of merging BBH, BNS, and BHNS systems. From the simulations, GW signals can be extracted and used to develop models to analyse the detected data. Similarly, the output describing the ejected material can be used to model spectra and light curves for EM transients, e.g., \cite{Tanaka:2013ana,Tanaka:2013ixa,Kawaguchi:2016ana,Dietrich:2016fpt,Perego:2017wtu,Kawaguchi:2019nju}, that can be compared to observations. \par 

Due to their computational cost, NR simulations typically cover only a few tens of milliseconds after the merger~\cite{Hotokezaka:2012ze,Kawaguchi:2015bwa,Kyutoku:2015,Sekiguchi:2016bjd,Radice:2016dwd,Kiuchi:2017zzg}. However, the kilonova itself can last several days up to a few weeks.  Subsequently, most studies using radiative transfer codes or (semi-) analytic models to calculate kilonova light curves,  e.g., \cite{Kasen:2017sxr, Tanaka:2017qxj, Banerjee:2020, Kawaguchi:2020osi, Korobkin:2021, Bulla:2019muo}, assume naturally a homologous expansion, i.e., that the radial velocity of each ejecta element remains constant. As the ejecta expands, the density and thus the speed of sound decreases rapidly until it is only a tiny fraction of the expansion speed. Under these circumstances, the different parts of the ejecta can no longer ``communicate'' through pressure waves and consequently can no longer influence each other, i.e., they are out of sonic contact. This means that the velocity structure can no longer change and the movement is homologous, but this condition might not be met at the end of a typical NR simulation. A first effort to include hydrodynamic effects using a spherically symmetric Lagrangian radiation hydrodynamics code to calculate kilonova light curves from NR data showed different results with and without the assumption of homologous expansion~\cite{Wu:2021ibi}. Refs.~\cite{Kawaguchi:2020vbf,Kawaguchi:2022bub} examined the ejecta evolution of BNS merger from NR simulations on a fixed gravitational background assuming axisymmetry for homologous expansion. It was shown that only $0.1$\,days after the merger the deviation of the radial velocity distribution from the assumption of homologous expansion is smaller than $1$\,\%, i.e., a homologous expansion is only evident afterwards. We note that this study considered multiple ejecta components, in particular post-merger components, which might delay the homologous expansion phase. \par

Using three-dimensional NR simulations, we want to readdress the problem by investigating how this assumption affects the calculation of the kilonova light curves, focusing on the dynamical ejecta component. Dynamical ejecta from NS merger simulations have been hydrodynamically evolved to late times before \cite{Rosswog:2013kqa,Grossman:2013lqa}. While these earlier studies were not fully relativistic, their Lagrangian nature allowed to evolve the ejecta up to $100$~years after the merger. Long-term evolutions are harder in Eulerian approaches, but first steps have already been taken to perform seconds-long NR simulations, e.g., using Cowling approximation~\cite{Hayashi:2021oxy} or assuming axisymmetry~\cite{Fujibayashi:2020dvr}. \par 

The aim of this study is to adapt our NR code \textsc{bam}~\cite{Bruegmann:2006at,Thierfelder:2011yi} for such long-term evolutions of the dynamical ejecta component and to investigate the degree of homology of the expanding material.\footnote{We note that~\cite{Fernandez:2014bra, Fernandez:2016sbf} showed that the interaction of multiple ejecta components affect the ejecta profile and are needed for realistic kilonova models. Hence, in future studies, we plan to simulate secular ejecta that are emitted on longer time scales up to seconds after the merger, which cannot be considered with the present method.} In particular, we modify the grid structure of our simulations after the merger, i.e., we coarsen the resolution to reduce the computational costs and to allow for faster simulations. Additionally, the size of the grid is increased to track the outflowing material for a longer period. Since these changes cause the strong-field region to be insufficiently resolved, we apply the Cowling approximation, i.e., we freeze the spacetime. We probe our new method by simulating two BNS systems with different Equations of State (EOSs), SLy and H4. For the simulations we use the NR code \textsc{bam}~\cite{Bruegmann:2006at,Thierfelder:2011yi}. The results are then transferred to the radiative transfer code \textsc{possis} \cite{Bulla:2019muo} to calculate light curves and to analyse the properties of the kilonova.\par 

The article is structured as follows. In Sec.~\ref{sec:methods}, we discuss the techniques used in our simulations and we describe the implemented changes. In Sec.\,\ref{sec:simulation}, we present the results from both BNS setups. In particular, we study the impact of the different EOSs and investigate the homologous nature of the expansion. Furthermore, the light curves of the kilonova are modelled to determine the impact of the homologous expansion assumption in Sec.\,\ref{sec:kilonova}. We summarise the main aspects and give a short outlook in Sec.~\ref{sec:summary}. Unless otherwise specified, we employ dimensionless units with $G=c= \mathrm{M}_\odot =1$. Further, we apply a metric with signature $\left(-+++\right)$.

\section{Methods}
\label{sec:methods}

\subsection{Standard Compact Binary Evolution}

\subsubsection{Spacetime and Matter Evolution}

\textsc{bam} employs method-of-lines for the dynamical evolution of the gravitational field, where we apply a fourth-order Runge-Kutta scheme and a Courant-Friedrichs-Lewy (CFL) coefficient of 0.25. For our NR simulations, Einstein's field equations are written in $3 + 1$ form. We use the Z4c reformulation \citep{Hilditch:2012fp, Bernuzzi:2009ex}, together with the 1+log slicing~\citep{Bona:1994b} and a Gamma driver shift condition~\citep{Alcubierre:2002kk} to ensure a long-term, stable evolution. \par

We perform pure General Relativistic Hydrodynamic (GRHD) simulations. The state of the fluid is fully described by the primitive variables $\mathbf{w}$, which comprise the proper rest-mass density $\rho_0$, the internal energy density $\epsilon$, the pressure $p$, and the fluid velocity $v^i$. The evolution equations are derived from the energy-momentum conservation law and conservation of particles. Introducing the conservative variables $\mathbf{q}$, which are the conserved rest-mass density $D$, the momentum density $S_i$, and the internal energy density $\tau$ as seen by the Eulerian observer, the resulting evolution system is written in the form of a balance-law, i.e., $\partial_t {\bf q} + \partial_k {\bf f}^k({\bf q}) = {\bf s}({\bf q})$. The conservative variables $\mathbf{q}$ are related to the primitive variables $\mathbf{w}$ by:

\begin{equation}
    D = \rho W, \hspace{0.5cm} S_i = \rho h W^2 v_i, \hspace{0.5cm} \tau = \rho h  W^2 -p - D,
\end{equation}

\noindent with the Lorentz factor $W=(1- v_i v^i)^{-1/2}$ and the specific enthalpy $h = 1+ \epsilon + p/\rho_0$. For a detailed discussion and derivation of the evolution equations, we refer to \cite{font:2000}.\par 

In order to close the evolution system, an EOS with $p = p\left(\rho_0, \epsilon\right)$ is needed. For the performed BNS simulations, we used piecewise-polytropic fits of the SLy EOS~\citep{Douchin:2001sv} and the H4 EOS~\citep{Lackey:2006} following \cite{Read:2008iy}. Of these two EOSs, SLy is softer and H4 is stiffer. The zero-temperature EOSs are extended to include thermal effects by adding a thermal pressure $P_{\rm th} = \left(\Gamma_{\rm th} - 1\right)\rho_0 \epsilon_{\rm th}$, see \citep{Bauswein:2010dn}. For the presented BNS simulations, we set $\Gamma_{\rm th} = 1.75$.

\subsubsection{Vacuum and Low-Density Treatment}

The simulation of the vacuum region surrounding the system is numerically challenging. One reason is the reconstruction of conservative variables \textbf{q} to primitive variables \textbf{w} due to the presence of the rest-mass density $D$ in the denominator~\citep{Thierfelder:2011yi}. Hence, the standard approach is to fill the vacuum with a cold, low-density static artificial atmosphere. The atmosphere density $\rho_{\rm atm}$ is typically defined as a fraction $f_{\rm atm}$ of the initial central density $\rho_c$ of the NS as $\rho_{\rm atm} = f_{\rm atm} \rho_c$. As soon as the density of a grid cell falls below a density threshold $\rho_{\rm thr}$ defined as a fraction $f_{\rm thr}$ of the atmosphere density, say $\rho_{\rm thr} = f_{\rm thr} \rho_{\rm atm}$, it is set to the atmosphere value $\rho_{\rm atm}$. \par 

Because the density difference between the NS and the artificial atmosphere is several orders of magnitude, the dynamical impact is often claimed to be negligible. This may be true for properties connected to the bulk motion such as the GW emission or the timing of the merger, but the assumption certainly breaks for outflowing ejecta. Since the density continues to decrease as the ejecta expands, it could eventually fall below the threshold $\rho_{\rm thr}$ and set to $\rho_{\rm atm}$. Consequently, the artificial atmosphere potentially distorts the ejecta simulation and should be avoided in order to obtain reliable results. \par 

In \textsc{bam}, besides the artificial atmosphere method, the ``vacuum method'', introduced in \cite{Poudel2020}, is implemented. The idea of the ``vacuum method'' is to set all matter variables to zero if the pressure $p$ in the conservatives to primitives reconstruction cannot be found. The variables of a grid cell are also set to zero if quantities are not physical, e.g., if the density is negative with $D < 0$ or if the energy density $\epsilon$ is complex. Furthermore, the flux computation at the interface between matter cells and vacuum cells must be adjusted to achieve physical results. This method allows for simulations with ``real vacuum" and is the preferred choice in this work.

\subsubsection{Grid Structure}

A common challenge in numerical simulations is to sufficiently resolve different length scales: the strong-field region inside and close to the NSs, and the far-field region where we extract GWs. For this purpose \textsc{bam} uses an Adaptive Mesh Refinement (AMR) technique following the ``moving boxes" approach employing a hierarchy of cell-centered nested Cartesian boxes. The numerical grid comprises $L$ refinement levels from $l=0$ being the coarsest to $l=L-1$ being the finest level. Following a $2:1$ refinement strategy, the resolution increases by a factor of two on every level. Accordingly, the grid spacing $h_l$ on level $l$ is determined by $h_l = 2^{-l} h_0$ for a fixed spacing $h_0$ on the coarsest level. Two successive refinement levels are called the parent level for the coarser, larger level $l$ and the child level for the finer, smaller level $l+1$. \par 

Each refinement level consists of one or more Cartesian boxes with equal grid spacing. For the inner refinement levels with $l > l_m$, the boxes can move and adjust dynamically during the evolution to ensure that the NSs are always covered by the refinement box with the highest resolution. The Cartesian refinement boxes have a fixed number of grid points in each direction. There is a distinction between an outer box with $n$ grid points and an inner moving box with $n_m$ grid points. As the grid spacing decreases for higher levels, the numerical domain of level $l$ is generally larger compared to its child level $l+1$. To increase the numerical domain but maintain the finest resolution, additional coarser levels can be attached to the grid structure, which is exploited in the new implementation. \par 

The CFL coefficient sets an upper bound for the time step $\Delta t$ depending on the resolution $\Delta x$, to ensure the stability of the simulation, through the relation $\left| \Delta t/\Delta x \right| \leq a$, where the CFL coefficient $a$ depends on the characteristic velocity of the simulated system and the employed numerical method. Because each refinement level uses a different grid spacing $\Delta x$, each level has different upper limits for $\Delta t$. Theoretically, the smallest time step determined by the finest refinement level can be applied for all other levels. However, this increases noticeably the computational time and slows down the simulation. For this reason, the Berger-Oliger scheme~\citep{Berger:1984zza} is used in \textsc{bam}, see \cite{Brugmann:2008zz}. The basic idea of \textsc{bam}'s Berger-Oliger implementation is simple: given the $2:1$ refinement strategy, the child level $l+1$ performs two time steps with $\Delta t_{l+1} = 2^{-1} \Delta t_l$, while the parent level $l$ evolves only one step with $\Delta t_l$. When successive levels are aligned in time, restriction and prolongation steps are applied to match the evolution with the different resolutions at the refinement levels. \par 

Every refinement level requires boundary conditions which are typically set by physical or symmetry conditions. Higher refinement levels use so-called buffer regions that are populated by prolongation of data from the parent level to the child level whenever the two levels are aligned in time. We use six buffer points and perform linear interpolation in time to update the buffer region for the substeps of the child level, see \citep{Brugmann:2008zz}.\par

The prolongation step to fill the buffer region generally carries numerical truncation errors. For this reason, we apply a correction step that ensures flux conservation across refinement boundaries. This is referred to as conservative mesh refinement (CAMR) and follows the Berger-Colella scheme~\citep{Berger:1989a}. For details on the CAMR implementation in \textsc{bam}, we refer to \cite{Dietrich:2015iva}. 

\begin{figure*}[t!]
    \centering
    \includegraphics[width=\linewidth]{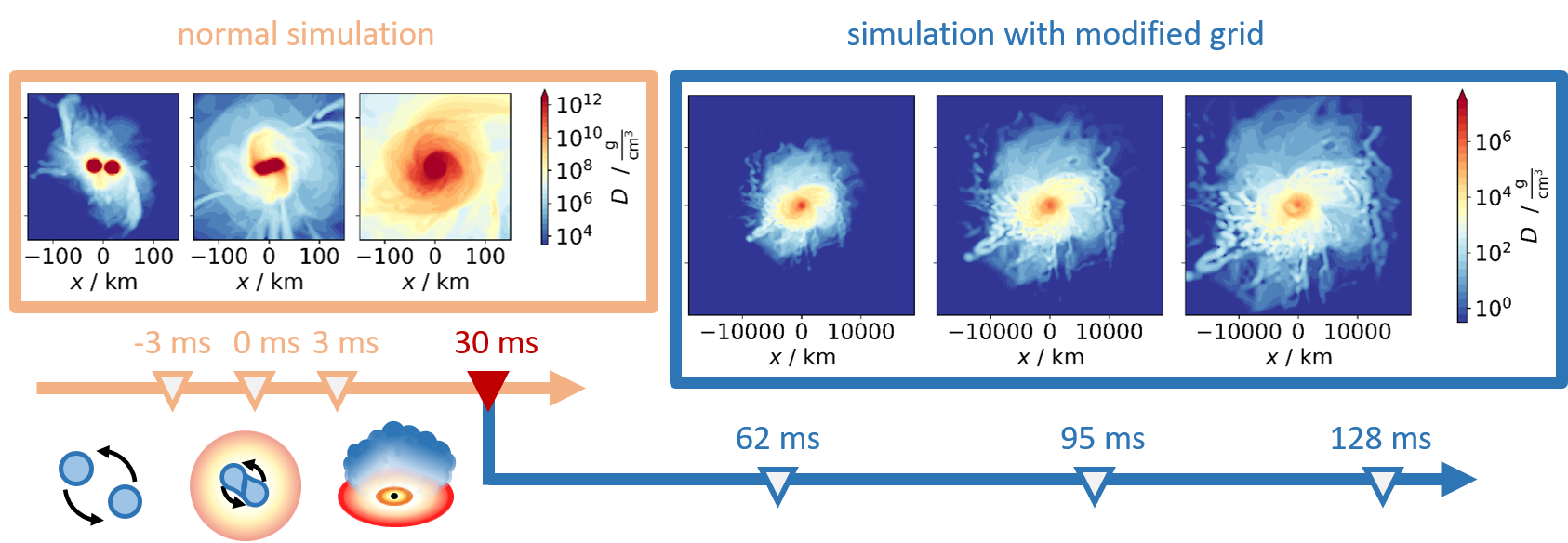}
    \caption{The timeline of a simulation. The orange arrow and orange frame represents the ``normal" simulation and the blue arrow and blue frame the ``modified" simulation with the Cowling approximation. The red triangle illustrates the checkpoint used for the grid modification. As example we show snapshots of the rest-mass density of the H4-128 and the H4-128-30ms simulation in the $x$-$y$ (the orbital) plane, see Tab.~\ref{tab:simulation_grid} and Tab.~\ref{tab:mod_paramter}. The given time of the snapshots are relative to the merger time.}
    \label{fig:checkpoints}
\end{figure*}

\subsection{Introducing a New Grid Structure}

Because NR simulations are computationally expensive, usually only a few tens or hundreds of milliseconds around the merger are covered. The longest NR simulations to date cover a few seconds, but are restricted to Cowling approximation~\cite{Hayashi:2021oxy} and axisymmetry~\cite{Fujibayashi:2020dvr}. Long-term simulations are essentially needed for a more comprehensive study of the ejected material and for a consistent understanding of the merger and post-merger processes. For this purpose, the computational costs must be reduced, which can be achieved with a lower grid resolution. But, since the simulation of the merger requires a well-resolved strong-field region, the resolution can only be reduced afterwards. Once the resolution is reduced, we use the Cowling approximation, i.e., we stop the evolution of the gravitational field and the spacetime is ``frozen in time".\par

Fig.~\ref{fig:checkpoints} shows the time sequence of the simulations for one example. We start with a simulation using our standard grid structure. In the top row, we show the merging process in one of our simulations, which takes only a few milliseconds. The formation of a stable remnant system, here a BH with accretion disk, requires a few tens of milliseconds. Modifying the existing checkpoint algorithm\footnote{The general purpose of the checkpoint algorithm is it to enable a restart of an existing simulation. As NR simulations can take several weeks or months, a running simulation may be aborted by processor problems or simply by limited walltime on a High Performance Computing system. For this reason, regular checkpoints are saved containing all the information about the grid, spacetime, and fluid variables at a given time. With the checkpoint the simulation can be continued from this time step.}, we use a written checkpoint after the collapse to change the grid structure and continue the ``modified" simulation with frozen spacetime and reduced resolution to allow for faster computation. 

\begin{figure}[t!]
    \centering
    \includegraphics[width=\linewidth]{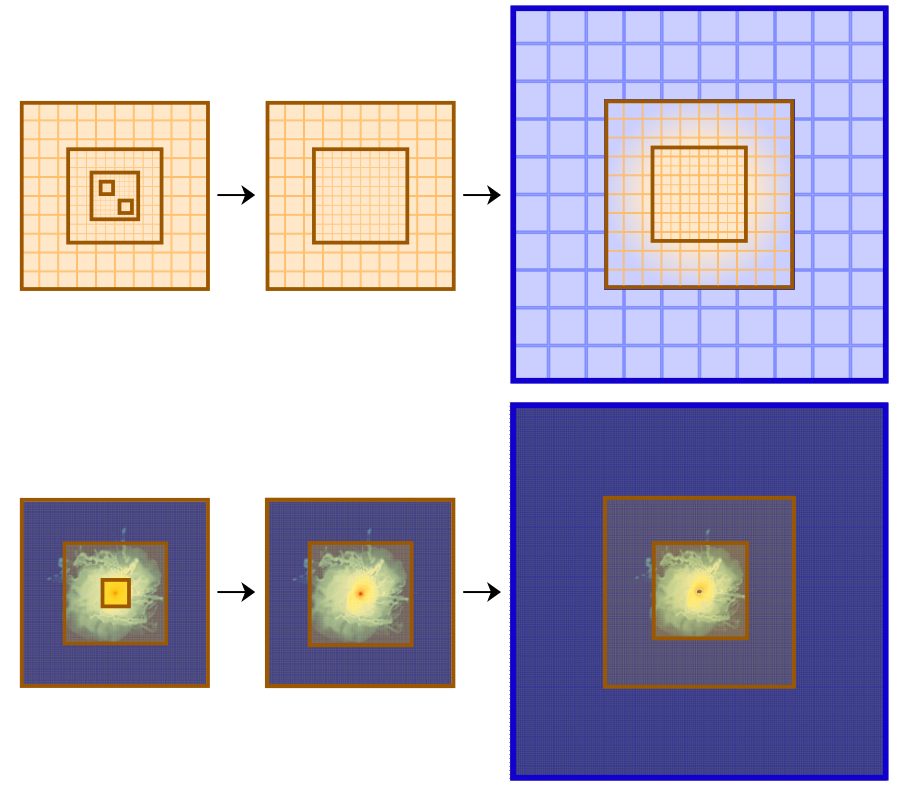}
    \caption{Modification of the grid structure to enable long-term simulations. In the top row, the grid itself is visualised: in yellow the data of the initial simulation and in blue the Schwarzschild spacetime for the extended region. In the bottom row, the rest-mass density $D$ for the grid change of one simulation is shown in the $x$-$y$ (the orbital) plane. The bound matter is removed.}
    \label{fig:newgrid}
\end{figure}

In Fig.~\ref{fig:newgrid}, the grid modification is illustrated in two dimensions. The original grid consists of several nested refinement levels including inner moving boxes. The first step to reduce the resolution is it to remove the finest refinement levels. The number of removed levels, $l_\textnormal{rm}$, determines the coarseness of the posterior simulation. The next step is it to extend the numerical domain by adding $l_\textnormal{add}$ new coarser refinement levels. In total, the modified grid structure comprises $L-l_\textnormal{rm}+l_\textnormal{add}$ refinement levels. For consistency, the initial level labels are shifted by $l_\textnormal{add}$, i.e., $l=0$ becomes $l=l_\textnormal{add}$, $l=1$ becomes $l=l_\textnormal{add}+1$, $l=2$ becomes $l=l_\textnormal{add}+2$, etc., and the new coarsest level starts again at $l=0$.\par

For the extended region, we assume Schwarzschild spacetime~\citep{Schwarzschild:1916}. For the initialisation, we use the remnant mass $M_{\rm rem}$ as an additional input parameter in the code. To ensure a smooth transition between the Schwarzschild spacetime in the outer regions and the original spacetime, a Planck Taper window function $f_\textnormal{PT}\left(r\right)$ is applied, which is set to $f_\textnormal{PT}\left(r\right)=0$ for $r\geq R_2$ and to $f_\textnormal{PT}\left(r\right)=1$ for $r< R_1$. For the transition region between $R_1 < r < R_2$, we use:

\begin{equation}    f_\textnormal{PT}\left(r\right) = \frac{1}{1+\exp\left(\frac{R_2-R_1}{R_2-r}-\frac{R_1-R_2}{R_1-r}\right)}.
    \label{eq:PT}
\end{equation}

\noindent Concretely, we multiply our original spacetime data by this function $f_\textnormal{PT}$ and the Schwarzschild metric by $\left(1-f_\textnormal{PT}\right)$ to have a smooth transition. Thus, the parameter $R_1$ defines the spatial range for which we keep the original spacetime data and $R_2$ defines the distance from which pure Schwarzschild spacetime is assumed. \par

Additionally, we implement a mask to distinguish grid cells containing bound and unbound matter when the grid structure is changed. We define unbound matter through: 
\begin{equation}
     u_0 < -1 \hspace{0.5cm} \mathrm{and} \hspace{0.5cm} v_r > 0.
    \label{eq:unbound}
\end{equation}

\noindent The first condition, the geodesic criterion, refers to the time component of the four velocity $u_0$ and requires an unbound trajectory of the fluid element, provided it follows a geodesic. The second condition demands an outward pointing radial velocity $v_r$. We denote the conserved rest-mass density for unbound matter by $D_u$. \par 

The effects of the grid modifications, are most severe in the dense, bound matter region around the remnant. Fluctuations in the metric are strongest here and may still affect the behaviour of the matter. In fact, the freezing of the metric and the reduced resolution causes some bound matter around the remnant to expand and become unbound, which can distort the results. For this reason, we remove this part from the simulation when the grid is changed. To verify that the evolution of the unbound matter is not affected by the modifications, we have run the ``normal" simulations alongside the ``modified" simulations a bit further. The comparison showed that the results are qualitatively the same.

\section{Binary neutron stars simulations}
\label{sec:simulation}

\begin{table}[t!]
    \centering
    \caption{Grid parameter of the ``normal'' simulations, from left to right: Simulation name, the total number of refinement levels $L$, the finest non-moving level $l_m$, the number of grid points in each direction for fixed boxes $n$ and for moving boxes $n_m$, the grid spacing on the coarsest level $h_0$ and on the finest level $h_{L-1}$ given in M$_\odot$, and the applied EOS.}
    \label{tab:simulation_grid}
    \begin{tabular}{l|c c c c c c c c}
    \hline
    Simulation & $L$   & $l_m$ & $n$ & $n_m$ & $h_0$ & $h_{L-1}$ & EOS \\ 
    \hline
    \hline
    H4-096          & 9     & 5 & 128     & 96    & 120   & 0.234     & H4 \\ 
    \hline 
    H4-128          & 9     & 5 & 170     & 128   & 90    & 0.176     & H4 \\ 
    \hline
    H4-144          & 9     & 5 & 192       & 144   & 80    & 0.156     & H4 \\ 
    \hline
    \hline
    SLy-096          & 9     & 5 & 128      & 96    & 120   & 0.234     & SLy \\ 
    \hline 
    SLy-128          & 9     & 5 & 170      & 128   & 90    & 0.176     & SLy \\ 
    \hline
    SLy-144          & 9     & 5 & 192      & 144   & 80    & 0.156     & SLy \\
    \hline
    \end{tabular}
\end{table}

\subsection{Configurations}

We construct initial data for two equal-mass BNS simulations both with gravitational masses $m_A = m_B = 1.35$\,M$_\odot$ using the pseudospectral code \texttt{SGRID} \citep{Tichy:2006qn,Tichy:2009yr}. We employ the SLy EOS and H4 EOS. The two NSs have baryonic masses of $m_{b,A} = m_{b,B} = 1.49$\,M$_\odot$ for SLy and $m_{b,A} = m_{b,B} = 1.47$\,M$_\odot$ for H4. We perform the simulations with three different resolutions: low, medium, and high with $96$, $128$, and $144$\,grid points covering the NS, see Tab.\,\ref{tab:simulation_grid}. Since our analysis focuses on the processes after the merger, we choose a small initial separation: $46.96$\,km, for the simulations with SLy and $37.70$\,km, for the simulations with H4. The two NSs merge already after two orbits at $t_\mathrm{merger} \approx 4.7$\,ms for the simulations with H4, and after seven orbits at $t_\mathrm{merger} \approx 19.7$\,ms for the simulation using the SLy EOS.

\begin{table}[t!]
    \centering
    \caption{Parameter to determine the grid changes of the ``modified'' simulations, from left to right: Simulation name (consisting of the name of the corresponding ``normal'' simulation and the time of the grid modification relative to the merger time in milliseconds), number of removed refinement levels $l_{\rm rm}$, number of added refinement levels $l_{\rm add}$, remnant mass $M_{\rm rem}$ to set Schwarzschild spacetime, $R_2$ and $R_1$ to specify the transition region with the Planck Taper window function. Further, we list the ejecta mass $M_{\rm ej}$. }
    \label{tab:mod_paramter}
    \begin{tabular}{l|c c c c c|c}
    \hline 
    Simulation  & $l_{\rm rm}$   & $l_{\rm add}$ & $M_{\rm rem}$   & $R_2$ & $R_1$ & $M_{\rm ej}$ \\
     & & & $\left[{\rm M}_\odot\right]$ & $\left[{\rm km}\right]$ & $\left[{\rm km}\right]$ & $\left[{\rm M}_\odot\right]$ \\
    \hline \hline
    H4-096-30ms   & 6     & 3   & 2.64    & 1916     & 1342 & 0.0035 \\ \hline
    H4-128-30ms   & 6     & 3   & 2.65    & 2211     & 1548 & 0.0022 \\ \hline
    H4-128-34ms   & 6     & 3   & 2.65    & 2211     & 1548 & 0.0021 \\ \hline
    H4-128-39ms   & 6     & 3   & 2.64     & 2211     & 1548 & 0.0022 \\ \hline
    H4-144-35ms   & 6     & 3   & 2.65     & 2211     & 1548 & 0.0030 \\ \hline
    \hline
    SLy-096-26ms   & 6     & 3   & 2.62     & 2211     & 1548 & 0.0141 \\ \hline
    SLy-128-25ms   & 6     & 3   & 2.63     & 2948     & 2064 & 0.0131 \\ \hline
    SLy-144-25ms   & 6     & 3   & 2.60     & 2948     & 2064 & 0.0178 \\
    \hline
    \end{tabular}
\end{table}

\begin{figure*}[t!]
    \centering
    \includegraphics[width=\linewidth]{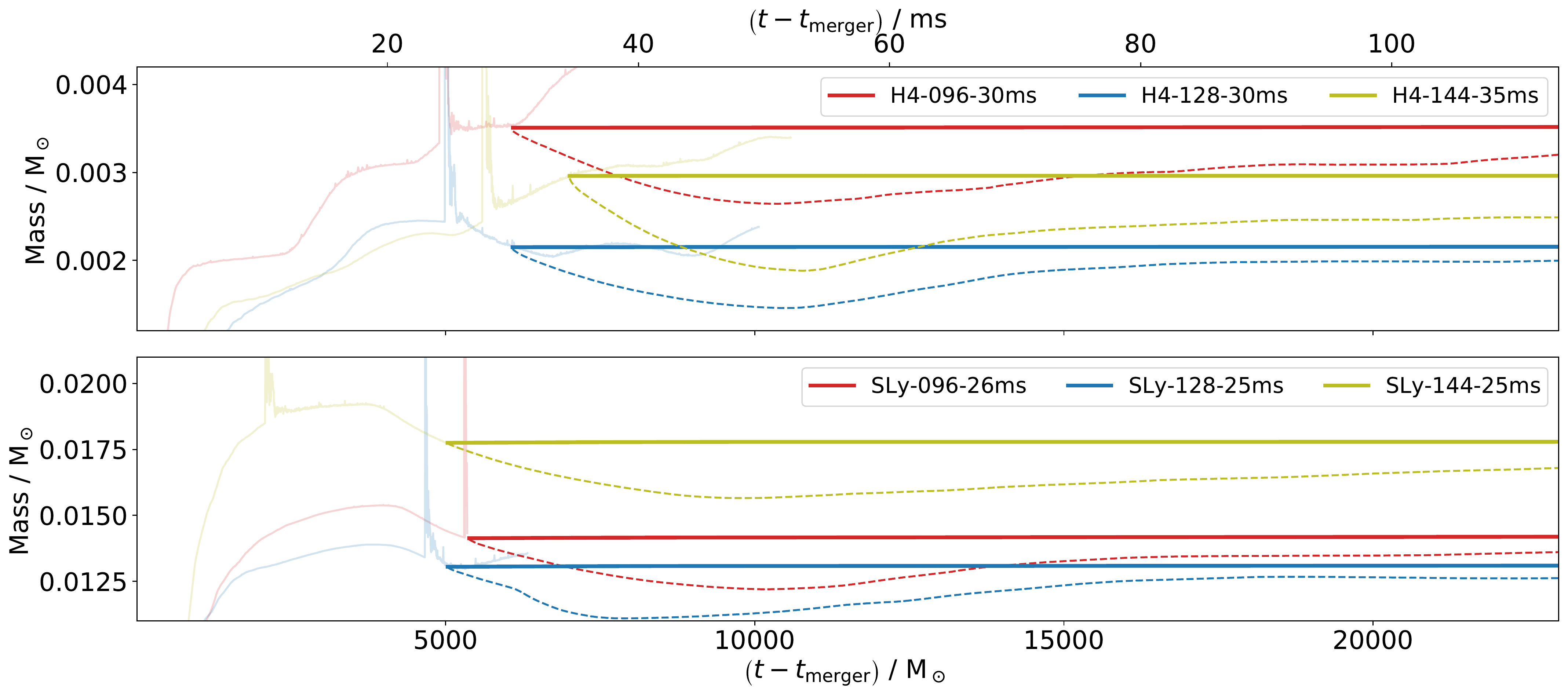}
    \caption{Evolution of the total rest-mass when bound matter is removed (solid lines) and the ejecta mass (dashed lines) in the simulation for low (red), medium (blue), and high (green) resolutions. As comparison, the ejecta masses of the ``normal" simulations before the grid modification are shown in faint lines. Top panel shows the results for the simulations with H4, bottom panel shows the results for the simulations with SLy. The masses are extracted from refinement level $l=0$.}
    \label{fig:BNSmass_res}
\end{figure*}

In both cases, a hypermassive NS (HMNS)\footnote{A HMNS is defined as a NS that exceeds the maximum mass of a uniformly rotating NS supported by the EOS, see \cite{Baumgarte:1999cq}, and is avoiding collapse due to differential rotation.} forms. As the system is dynamically unstable, it usually forms a BH within a few tens of milliseconds. In the simulations with H4, the lifetime of the HMNS is $\tau_\mathrm{HMNS} \approx 25$\,ms, and in the simulations with SLy, it is $\tau_\mathrm{HMNS} \approx 16$\,ms. The time of the collapse varies for different resolutions: for the simulation with H4 by $\pm 2$\,ms and for the simulations with SLy by $\pm 7$\,ms. Generally, the results for $\tau_\mathrm{HMNS}$ are in agreement with \cite{Dietrich:2015iva}. \par 

The determination of the lifetime $\tau_\mathrm{HMNS}$ is crucial for the choice of an appropriate time for the grid change $t_{\rm ch}$. We assume a stationary spacetime after the collapse. Therefore, we chose the time for the grid modification to be $t_{\rm ch} > t_{\rm merger} + \tau_{\rm HMNS}$. For the simulations with H4 we use $t_{\rm ch}=30$\,ms after merger and for the simulations with SLy we use $t_{\rm ch}=25$\,ms after merger. Because, the HMNS lifetime for the H4-144 simulation is slightly longer, a later change time is chosen with $t_{\rm ch} = 35$\,ms after merger. The same applies for the SLy-096 simulation, for which we take $t_{\rm ch} = 26$\,ms. Furthermore, we select two additional change times for the H4-128 simulation to determine the influence of $t_{\rm ch}$.\par

The parameters for changing the grid, such as the number of removed $l_{\rm rm}$ and added $l_{\rm add}$ refinement levels, are listed in Tab.~\ref{tab:mod_paramter}. The bound mass is removed from the simulation when the grid is changed and thus only the dynamical ejecta is evolved in the posterior simulation. With the employed changes, the ``modified'' simulations after grid change are about $6$ times faster than the ``normal'' ones.

\subsection{Dynamical Ejecta}

The dynamics of the post-merger relates strongly with the binary parameters, such as the total mass, the mass ratio, and the spins of the NSs, see, e.g., \citep{Nedora:2020qtd}. We focus here on the effects of the different EOSs and discuss the influences of different resolutions. There are, broadly speaking, two important mechanisms that produce the dynamical ejecta: heating due to shocks at the collision interface and core bounces, and the torques of the system causing tidal ejecta. For our equal mass BNS merger with SLy, i.e., the softer EOS, we find that the shock heating is more dominant than for the H4 setups, see, e.g., \cite{Sekiguchi:2016bjd}.\par 

In Fig.~\ref{fig:BNSmass_res}, the evolution of the ejecta masses and the total rest-masses for each resolution are compared. The upper panel shows the results of the simulations with H4 and the lower panel shows the results of the simulations with SLy. The ejecta mass of the simulations using SLy is significantly larger with $M_{\rm ej} \approx1.5\times 10^{-2}$\,M$_\odot$ than the ejecta mass of the simulations using H4 with $M_{\rm ej} \approx 2\times 10^{-3}$\,M$_\odot$. Previous studies also showed for equal mass BNS mergers, that the ejecta can be larger for softer EOSs than for stiffer EOSs, e.g., \cite{Hotokezaka:2012ze, Bauswein:2013yna, Dietrich:2015iva, Sekiguchi:2016bjd,rosswog22b}. The physical reason is that the stars are more compact for softer EOSs and merge with greater velocities at smaller orbital distances, which makes the encounters more violent. \par 

Because the bound matter is removed when the grid is changed, the total rest-mass in the ``modified" simulations coincides with the ejecta mass of the ``normal" simulation at $t_{\rm ch}$. For the simulations with H4 the difference in ejecta mass between the medium and low resolution is $1.3 \times 10^{-3}$\,M$_\odot$ which is by a factor of $1.625$ larger compared to the difference between the high and medium resolution with $0.8 \times 10^{-3}$\,M$_\odot$. Also for the simulations with SLy the ejecta mass varies for different resolutions: low and medium resolution differ by $1.0 \times 10^{-3}$\,M$_\odot$ and medium and high by $4.7 \times 10^{-3}$\,M$_\odot$.\par

The ``modified" simulations for H4 as well as for SLy show almost perfect conservation of the total rest-mass. However, the ejecta mass is not constant in the ``modified" simulations. This is not surprising and can be explained by the following considerations. On the one hand, the Cowling approximation and the assumption of Schwarzschild spacetime at large distances compromise the geodesic criterion, see Eq.~\eqref{eq:unbound}, and thus the determination of the unbound matter. On the other hand, the removal of matter at the centre leads to a lack of pressure. In fact, part of the matter falls back and no longer fulfils the second condition of Eq.~\eqref{eq:unbound}. As a consequence, the ejecta mass decreases initially. We observe this drop in the ejecta mass until $\Delta t \approx 25$\,ms after the grid change for all simulations. For this reason, we use the total rest-mass density $D$ instead of $D_u$ in the following analysis of ejecta evolution. Since we removed the bound part of the matter in the ``modified'' simulation, this corresponds to the dynamical ejecta component. \par 

\begin{figure}[t!]
    \centering
    \includegraphics[width=\linewidth]{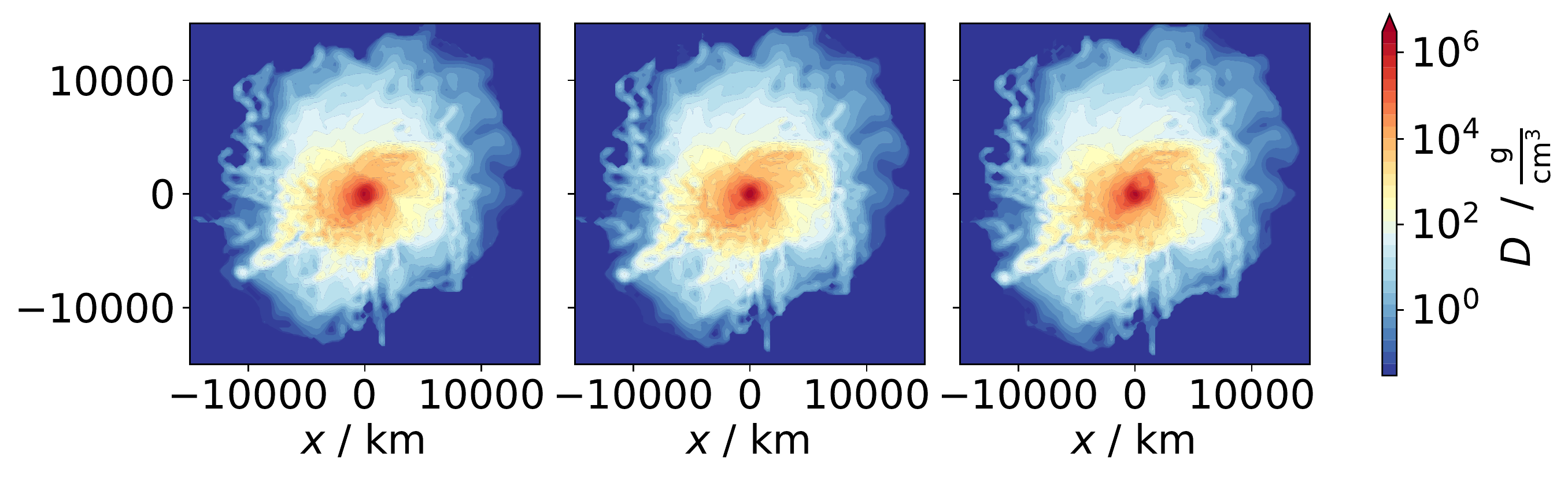}
    \hfill
    \caption{Snapshot of the rest-mass density $D$ in the orbital plane at $t=59$\,ms after the merger. From left to right: H4-128-30ms, H4-128-34ms, and H4-128-39ms.}
    \label{fig:BNSsnaps_changetimes}
\end{figure}

To ensure that our results are independent of the time of the grid change, we used for the H4-128 simulation three different times $t_{\rm ch}$ for the grid modification. We show two-dimensional plots of the rest-mass density $D$ in Fig.~\ref{fig:BNSsnaps_changetimes} to compare the qualitative differences in the evolution. The snapshots show the distributed ejecta in the $x$-$y$ plane at $t=59$\,ms after the merger. The images show almost identical results for different change times $t_{\rm ch}$, i.e., the behaviour is consistent and independent of the time for the grid change.

\subsection{Expansion of the Ejecta}

\begin{figure*}[t!]
    \centering
    \includegraphics[width=\linewidth]{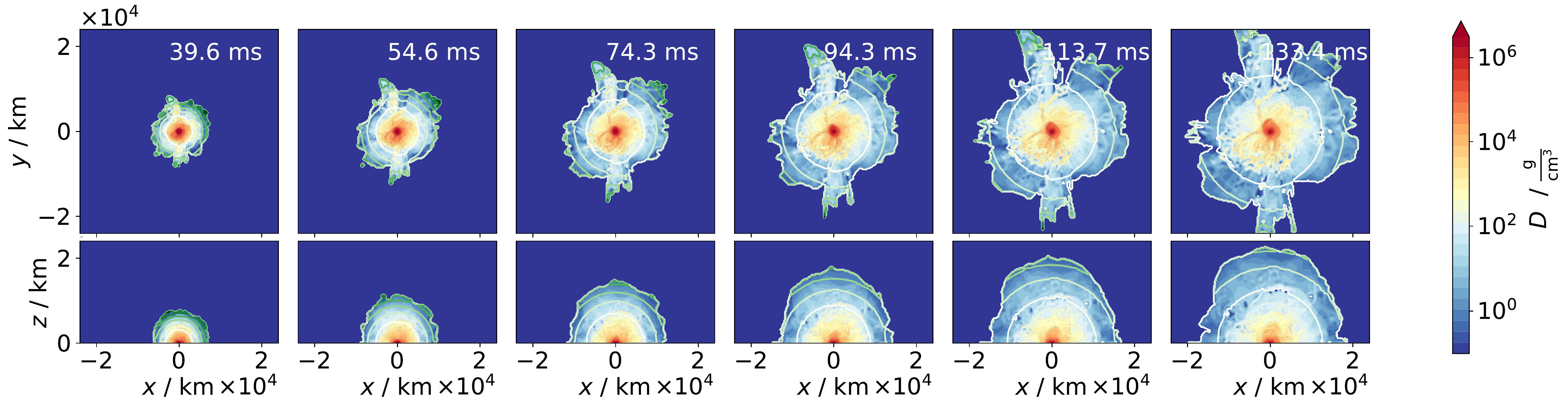}
    {(a) Snapshots of H4-144-35ms.}
    \hfill
    \includegraphics[width=\linewidth]{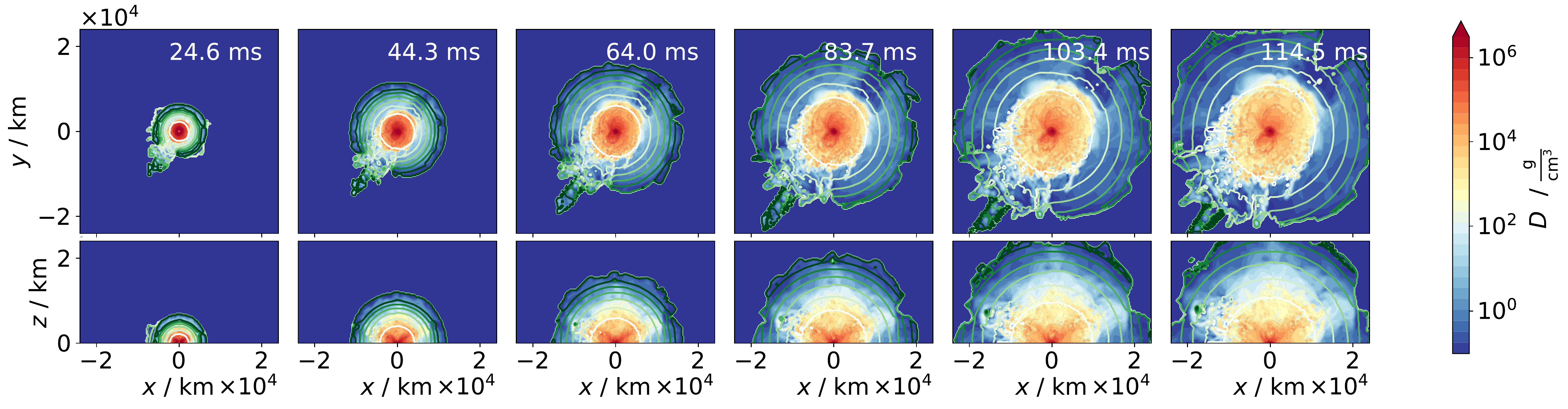}
    {(b) Snapshots of SLy-144-25ms.}
    \caption{The evolution of the outflowing material. The snapshots show the rest-mass density $D$ in $x$-$y$ and $x$-$z$ plane on refinement level $l=2$ for the H4-144-35ms and SLy-144-25ms simulation at six different times after the merger. The radial velocities $v_r$ for ($0.1$, $0.2$, $0.3$, $0.4$, $0.5$, $0.6$)\,$c$ are plotted as contour lines from white to dark green.}
    \label{fig:BNSsnaps_expansion}
\end{figure*}

We study the expansion of the dynamical ejecta component by analysing the time evolution of the rest-mass density distribution. Fig.~\ref{fig:BNSsnaps_expansion} shows snapshots of the rest-mass density $D$ in the $x$-$y$ and in the $x$-$z$ plane for different times after the merger. In addition, contour lines for the distribution of radial velocities are plotted from $v_r = 0.1$\,$c$ (white line) to $v_r=0.6$\,$c$ (dark green line) in $\Delta v_r = 0.1$\,$c$ steps.

For both systems, the overall distribution appears to be fairly spherical. Accordingly, shock heating seems to be the primary source of the dynamical ejecta. If tidal disruption had been more dominant, the tidal force would distribute the ejecta in the orbital plane resulting in a spheroidal distribution. There are deviations from the spherical distribution. In particular, the negative $x$ and negative $y$ quarters of the SLy plots show a fissured structure. This material is already ejected at the beginning of the simulation by artificial shocks at the surface of the NSs. Since we later choose the azimuth $\Phi = 0 $ for the calculation of the light curves, i.e., along the positive x-axis, these numerical errors should not affect our final results.\par

The velocity fronts maintain a spherical shape throughout the entire simulation, which is expected for a homologous expansion. The consistency of the overall structure is also an essential feature. For a more detailed analysis of the expansion, we compute the ejecta mass $m_r$ inside a sphere with radius $r$ via:

\begin{equation}
    m_{r} := \int_0^r \int_0^{2\pi} \int_0^{\pi}  D  \sin{\Theta} dr'd\Phi' d\Theta'. 
    \label{eq:mr}
\end{equation}

\begin{figure}[t!]
    \centering       
    \includegraphics[width=\linewidth]{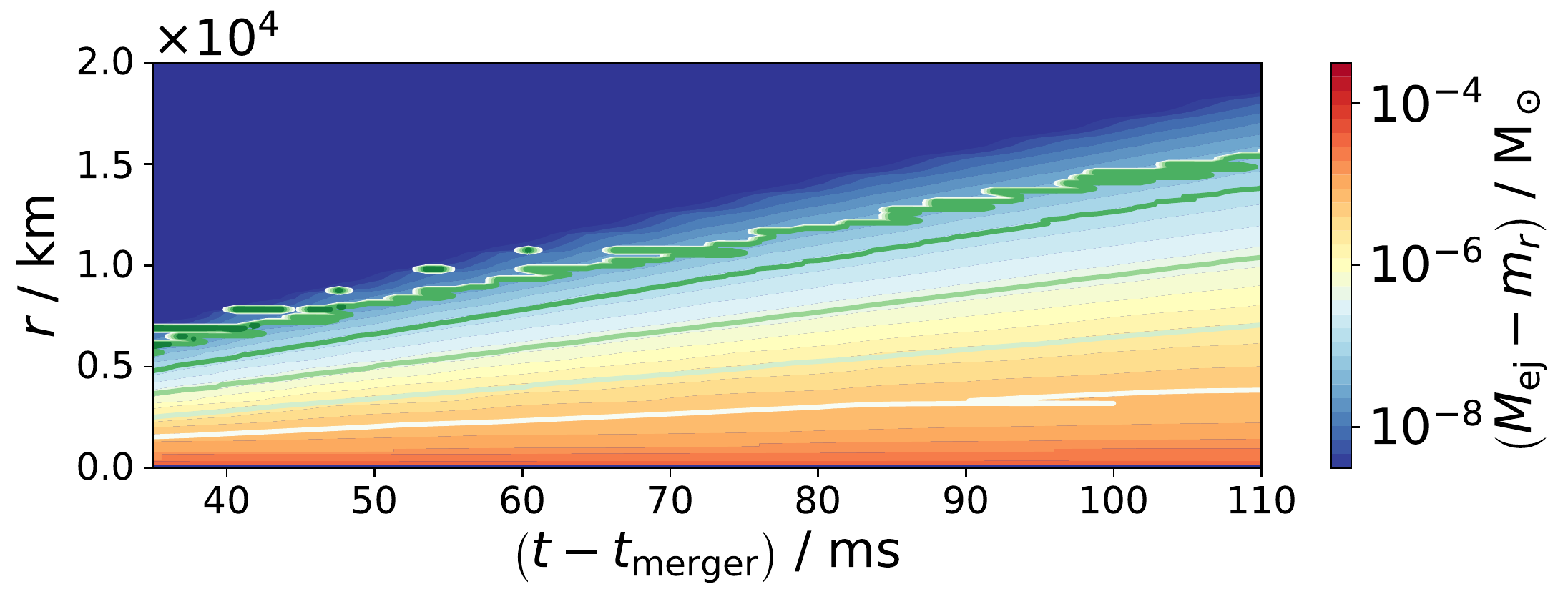}
    \vspace{0.2cm}
    {(a) H4-144-08025M.}
    \hfill
    \includegraphics[width=\linewidth]{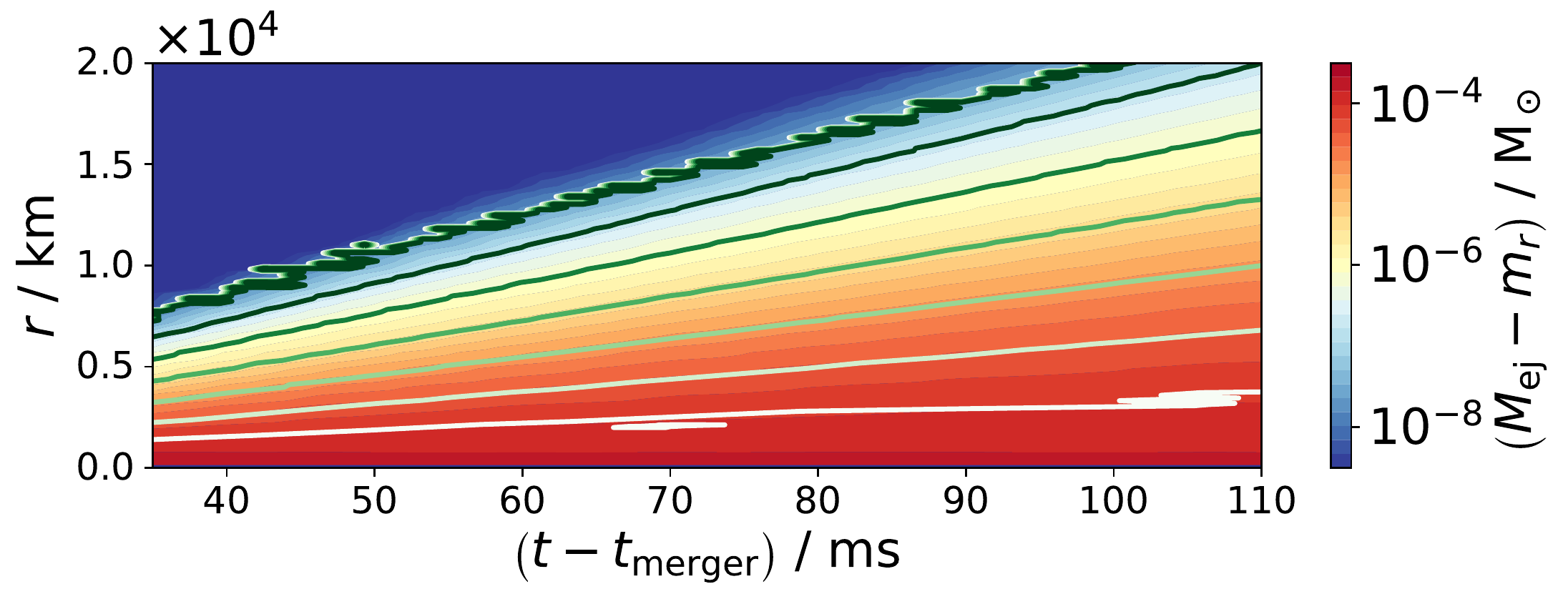}
    \vspace{0.2cm}
    {(b) SLy-144-25ms.}
    \caption{Analysing the expansion of the ejecta by considering the evolution of the masses $m_r$ within a sphere with radius $r$. The quantities are extracted from refinement level $l=2$. The contour lines in white to green indicate the mean radial velocities $\Bar{v_r}$ for ($0.1$, $0.2$, $0.3$, $0.4$, $0.5$, $0.6$)\,$c$.}
    \label{fig:BNSexpansion}
\end{figure}

\noindent When the ejecta expands, the radius $r$ of the sphere containing $m_r$ increases. In the case of homologous expansion, the radius should increase linearly. The mass spheres are traced by considering $m_r$ as a function of $r$ evolved in time. However, as discussed above, the evolution of the ejecta in the central region might be biased by the implemented modifications. Therefore, we consider instead the mass outside the sphere with $\left(M_{\rm ej}-m_r\right)$. The results are shown in Fig.~\ref{fig:BNSexpansion}. The evolution of the radius for the mass spheres in time is clearly visible. In fact, we find an almost linear dependence, indicating a homologous expansion of the dynamical ejecta in both simulations.

In addition, we compute the mean radial velocities $\Bar{v}_r$ for each shell of mean radius $r$. The radial profiles of the velocity $\Bar{v}_r$  is included in Fig.~\ref{fig:BNSexpansion} by contour lines from $\Bar{v}_r =0.1$\,$c$ (white line) to $\Bar{v}_r=0.6$\,$c$ (dark green line) in $\Delta \Bar{v}_r = 0.1$\,$c$ steps. The contour lines of the radial velocity are almost perfectly linear and agree well with the expansion of the mass spheres in both systems. Thus, our analysis indicates that homologous expansion is reached during our simulation. \\

\begin{figure*}[t!]
    \centering
    \includegraphics[width=\linewidth]{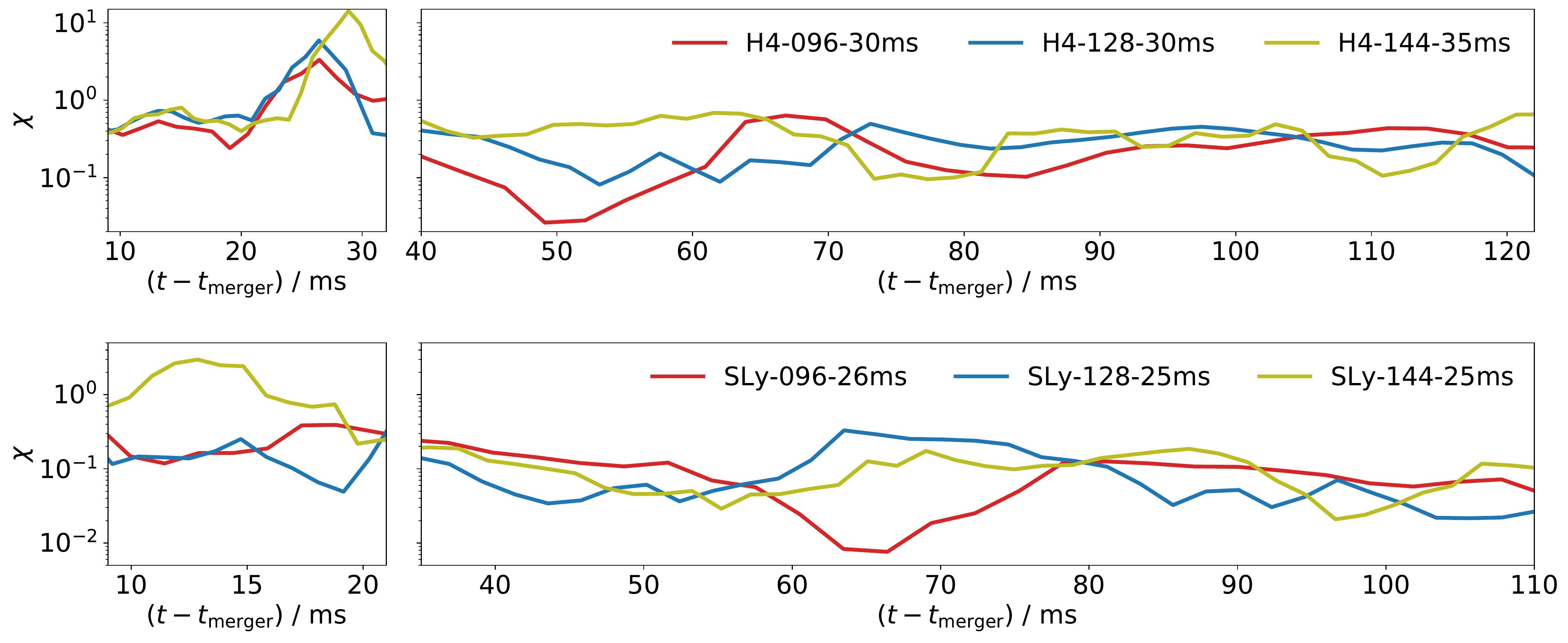}
    \caption{The homology parameter $\chi$, Eq.~\eqref{eq:homology}, as a function of time after the merger for all simulations. The left panels show the parameter for the ``normal" simulations extracted from refinement level $l=0$, and the right panels show the parameter for the ``modified" simulations extracted from refinement level $l=1$. }
    \label{fig:homology}
\end{figure*}

For a more quantitative investigation, we use the approach of \cite{Rosswog:2013kqa} and define a homology parameter: 
\begin{equation}
    \chi := \frac{\Bar{a}t}{\Bar{v}},
    \label{eq:homology}
\end{equation}

\noindent with average acceleration $\Bar{a}$ and average velocity $\Bar{v}$ of the dynamical ejecta. The homology parameter $\chi$ specifies whether the expansion is accelerated or homologous, i.e., for a constant acceleration $\chi \longrightarrow 1$ and for a constant velocity $\chi \longrightarrow 0$. The results for $\chi \left(t\right)$ are summarized in Fig.~\ref{fig:homology}. Overall, the values for $\chi$ are higher before the grid change and lower afterwards. More precisely, after the grid modification, the parameter is $\lesssim 0.5$ in the simulations with H4 and $\lesssim 0.2$ in the simulations with SLy. In particular, at around $100$\,ms after merger, the expansion deviates by $\sim (10 - 30)$\,\% from a perfect homologous expansion. 

The homology parameter in \cite{Rosswog:2013kqa} is generally smaller and has values of about $10^{-2}$ at $100$\,ms after the merger. A difference is also that whereas the parameter tends to decrease in our simulation, $\chi$ initially increases in \cite{Rosswog:2013kqa} and reaches a maximum of $\sim 10^{-1}$ after one second; cf. Fig.~$7$ in \cite{Rosswog:2013kqa}. This is because the latter work also implemented the nuclear heating from r-process according to \cite{Korobkin:2012uy} which continuously injects thermal energy into the ejecta and thus delays reaching the homologous phase.

\section{Kilonova Properties}
\label{sec:kilonova}

We use the three-dimensional Monte Carlo radiative transfer code \textsc{possis}~\cite{Bulla:2019muo} to model kilonova light curves based on our ejecta simulations (see Appendix~\ref{app:possis}). \textsc{possis} requires input data of the ejecta including the density, velocity, and electron fraction at a reference time $t_{0}$ to calculate kilonova light curves. Subsequently, the grid is evolved for each time step $t_j$ assuming homologous expansion. The velocity $\Vec{v}_i$ of each fluid cell $i$ remains constant, while the grid coordinates evolve following a homologous expansion. \par 

To probe how the deviations from a perfect homologous expansion influence the computation of the light curves, we extract the ejecta quantities at six different times after the grid change for each simulation. The snapshots in Fig.\,\ref{fig:BNSsnaps_expansion} represent the six reference times $t_0$ for which the ejecta data is extracted to start the radiative transfer simulations. If the assumption of homologous expansion is correct, the results should be independent of the extraction time $t_0$.

\begin{figure*}[t!]
    \centering
    \includegraphics[width=\linewidth]{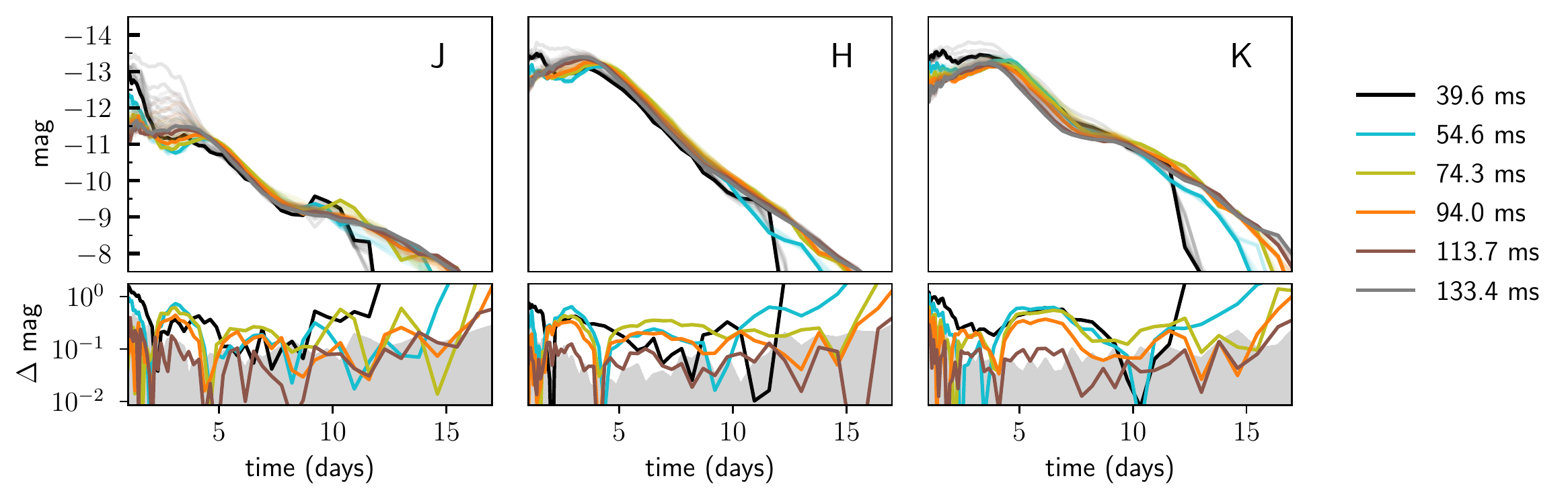}
    \caption{Top panels: Light curves for the H4-144-35ms simulation. The input data are extracted from refinement level $l=2$. The labels give the reference time $t_0$ in milliseconds after the merger for modelling the light curves. We compute the light curves for the azimuth $\Phi = 0$. While the results are shown for all viewing angles $\theta_{ \rm obs}$, the light curves for $\theta_{\rm obs} \neq 0$ are shown as faint lines. Bottom panels: Differences of the light curves relative to the curve of $t_0 = 133.4$\,ms. To compare with the Monte Carlo noise, we have calculated the light curve for $t_0 = 133.4$\,ms a total of six times and plotted twice the maximum difference from the mean value as a grey shadow in the background.}
    \label{fig:lightcurvesH4144}
\end{figure*}

\begin{figure*}[t!]
    \centering
    \includegraphics[width=\linewidth]{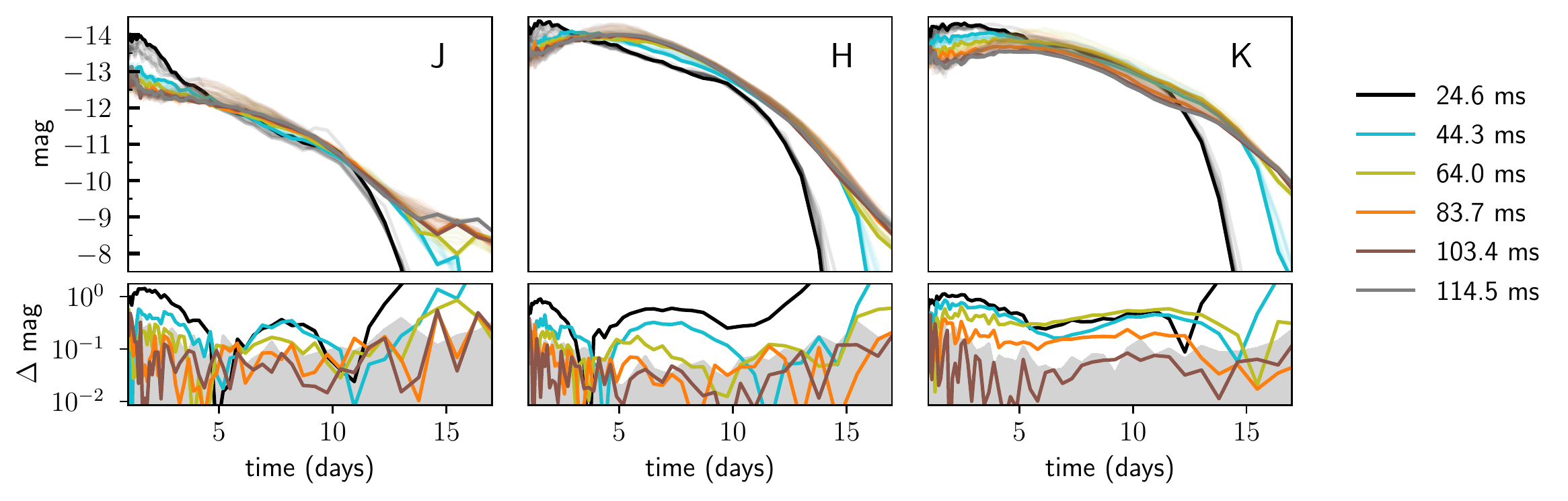}
    \caption{Top panels: Light curves for the SLy-144-25ms simulation. The input data are extracted from refinement level $l=2$. The labels give the reference time $t_0$ in milliseconds after the merger for modelling the light curves. We compute the light curves for the azimuth $\Phi = 0$. While the results are shown for all viewing angles $\theta_{ \rm obs}$, the light curves for $\theta_{\rm obs} \neq 0$ are shown as faint lines. Bottom panels: Differences of the light curves relative to the curve of $t_0 = 114.5$\,ms. To compare with the Monte Carlo noise, we have calculated the light curve for $t_0 = 114.5$\,ms a total of six times and plotted twice the maximum difference from the mean value as a grey shadow in the background.}
    \label{fig:lightcurvesSLy144}
\end{figure*}

We note that at the time when our work started, \textsc{bam} could not evolve the electron fraction $Y_e$ (see \cite{Gieg:2022mut} for the implementation of the $Y_e$ evolution in \textsc{bam}), which is why we have to make assumptions considering a shocked and an un-shocked component, that can be associated with a Lanthanide-free and a Lanthanide-rich component, respectively. Previous studies showed that both components are required to reproduce the kilonova observation AT2017gfo, i.e., to explain the early blue part of the light curve, and the long-term near-infrared emission \citep{Kasen:2017sxr}. We define an entropy indicator $\hat{S} = p/p\left(T=0\right)$. The entropy indicator $\hat{S}$ is high if the thermal component of the pressure $P_{\rm th}$ is large. Thus, for the ejecta caused by shock heating, $\hat{S}$ is expected to be higher than for the ejecta caused by torque. We set accordingly the electron fraction $Y_e$ lower for low $\hat{S}$ and higher for high $\hat{S}$. More precisely, using a threshold $\hat{S}_{\rm th} = 50$, the electron fraction of the grid cells with $\hat{S} > \hat{S}_{\rm th}$ is set to $Y_e = 0.3$ and for $\hat{S} < \hat{S}_{\rm th}$ to $Y_e = 0.15$.\par

We first calculated the light curves using $D_u$ as input. However, the determination of $D_u$, see Eq.\,\eqref{eq:unbound}, is impaired by the Cowling approximation and the implemented modifications, as discussed above, leading to an artificial decrease of the ejecta mass, see Fig.\,\ref{fig:BNSmass_res}. The light curves for the corresponding reference times are consequently less bright. To avoid this bias, we use instead the total rest-mass density $D$ of our simulations as input for \textsc{possis} in the presented results. Since we remove the bound matter with the grid modification, this total mass consists of the dynamical ejecta component only.\par 

Fig.~\ref{fig:lightcurvesH4144} and Fig.~\ref{fig:lightcurvesSLy144} show the results of the simulations with the highest resolution, i.e., H4-144-35ms and SLy-144-25ms. We focus on the infrared bands J, H, and K, since these are the dominant frequency bands for the simulated dynamical ejecta component. Shorter wavelengths in the optical bands, such as the u-, g-, r-, i- and z-bands, show similar behaviour but are less bright. \par

The light curves for the simulations with H4 show an earlier and sharper peak at about $\sim 4$\,days after the merger in the H and K bands. This is followed by a relatively steady decline. The peak of the light curves for the simulations with SLy is less sharp and rather flat, and is only reached about $\sim 5$ days in the H and K bands. Overall, the light curves for the simulations with SLy are brighter than for the simulations with H4, due to the higher ejecta mass. The differences for the viewing angles $\theta_{\rm obs}$ are small which can be explained by the small deviations from spherical symmetry of the ejecta input, see Fig.~\ref{fig:BNSsnaps_expansion}.\par

Our results show that the light curves for different extraction times are very consistent in each case. The light curves for $t_0 < 60$\,ms have an earlier and faster drop, but generally differ by only $\lesssim 1$\,mag until $\sim 12$\,days after the merger. For $t_0 > 80$\,ms the differences are mostly $\lesssim 0.4$\,mag and for $t_0 > 100$\,ms even within the Monte Carlo noise range. This shows that the assumption of a homologous expansion from $ t_0 > 80$\,ms after the merger seems to be absolutely valid.  

\begin{table}[t!]
    \centering
    \caption{A selection of NR simulations in the literature for which the ejecta data have been used in studies modelling kilonova light curves. Listed are references for BNS or BHNS simulations, and when/how the ejecta properties were extracted.}
    \label{tab:lc_review}
    \begin{tabular}{l|c r}
    \hline
    Reference                   & System & Extraction of Ejecta \\
    \hline
    \citet{Hotokezaka:2012ze}    & BNS   & at $t_0 \approx 10$\,ms after the merger \\
    \citet{Sekiguchi:2016bjd}    & BNS   & at $t_0 \approx 30$\,ms after the merger \\
    \citet{Kiuchi:2017zzg}       & BNS   & at $t_0 \approx 30$\,ms after the merger \\
    \citet{Radice:2016dwd}       & BNS   & on a sphere with $r \approx 295$\,km \\
    \citet{Kawaguchi:2015bwa}    & BHNS  & at $t_0 \approx 10$\,ms after the merger \\
    \citet{Kyutoku:2015}         & BHNS  & at $t_0 \approx 10$\,ms after the merger \\ \hline
    \end{tabular}
\end{table}

Most previous studies calculating kilonova light curves using radiative transfer codes or semi-analytical light curve models based on ejecta data from NR simulations use an idealised geometry. In this context, the light curve models of \cite{Tanaka:2013ana, Tanaka:2013ixa, Perego:2017wtu, Dietrich:2016fpt, Kawaguchi:2016ana, Kawaguchi:2019nju} utilize the NR simulations listed in Tab.~\ref{tab:lc_review}. In these, the ejecta is already extracted at $\left(10-30\right)$\,ms after merger. Also for \cite{Radice:2016dwd}, which uses an extraction sphere of $r \approx 295$\,km, an extraction time of $t_0 \approx 10$\,ms can be associated assuming a low ejecta velocity of $v = 0.1$\,$c$ and even earlier for higher velocities. Also in a recent study~\cite{Kullmann:2021gvo, Just:2021vzy} based on smoothed particle hydrodynamics homologous expansion is assumed at $\left(10-20\right)$\,ms after the merger. Our results show that the assumption of a homologous expansion at this time might bias the light curves, and that a later extraction time would be better.

\section{Conclusion}
\label{sec:summary}

In this article, we have presented a simple method to perform longer simulations of the dynamical ejecta with \textsc{bam}. By changing the grid structure of our code and applying the Cowling approximation after the collapse of the merger remnant and the formation of a BH, we were able to reduce the computational cost and speed up the simulation. This allowed us to perform long-term simulations of the ejecta in a reasonable computational time. We demonstrated our new method by simulating two equal mass BNS systems with different EOSs. With our new framework, the speed of the simulations increased by a factor of six.\par 

We used our simulations to test when homologous expansion of the ejecta is reached. Our results show that, although the expansion generally appears very homologous, deviations of around $\left(10 - 30\right)$\,\% from a perfectly homologous expansion are still present at $100$\,ms after the merger. To investigate how this affects the light curves, we used our data as input for radiative transfer simulations and modelled kilonova light curves. The results show that $\sim 80$\,ms after the merger the differences in the light curves are negligible. Thus, previous studies that used NR simulations and extracted ejecta properties already $(10 - 30)$\,ms after merger appear rather optimistic, as the expansion may not be fully homologous yet.\par 

While our results focus on the dynamical ejecta component and equal mass systems, additional work is needed for an accurate description of the ejecta evolution and kilonova light curves, in particular, through the inclusion of other ejecta components.

\begin{acknowledgments}
We thank P.~Biswas, B.~Br\"ugmann, M.~Emma, M.~Mattei, V.~Nedora, H.~Pfeiffer, and F.~Schianchi for helpful discussions.

M.B. acknowledges support from the Swedish Research Council (Reg. no. 2020-03330). S.V.C. was supported by the research environment grant ``Gravitational Radiation and Electromagnetic Astrophysical Transients (GREAT)'' funded by the Swedish Research council (VR) under Grant No. Dnr. 2016-06012. SR has been supported by the Swedish Research Council (VR) under grant number 2020-05044, by the research environment grant ``Gravitational Radiation and Electromagnetic Astrophysical Transients'' (GREAT) funded by the Swedish Research Council (VR) under Dnr 2016-06012, and by the Knut and Alice Wallenberg Foundation
under grant Dnr. KAW 2019.0112.

The simulations were performed on the national supercomputer HPE Apollo Hawk at the High Performance Computing (HPC) Center Stuttgart (HLRS) under the grant number GWanalysis/44189, on the GCS Supercomputer SuperMUC\_NG at the Leibniz Supercomputing Centre (LRZ) [project pn29ba], and on the HPC systems Lise/Emmy of the North German Supercomputing Alliance (HLRN) [project bbp00049]. 

\end{acknowledgments}

\appendix

\section{\textsc{possis}}
\label{app:possis}

\textsc{possis} is a Monte Carlo radiative transfer code~\cite{Bulla:2019muo} that requires input for a three-dimensional grid at a reference time $t_{0}$. The input data represent a snapshot of the ejecta. Subsequently, the grid is evolved for each time step $t_j$ assuming a homologous expansion. In particular, the velocity $\Vec{v}_i$ of each fluid cell $i$ remains constant, while the grid coordinates evolve. The density at time $t_j$ within the cells is determined by:

\vspace{-0.2cm}
\begin{equation}
    \rho_{ij} = \rho_{i,0} \left(\frac{t_j}{t_0}\right)^{-3},
\end{equation}

\noindent with the rest-mass density $\rho_{i,0}$ as initial density at $t_0$.\par 

The code generates photon packets at each time step that propagate through the ejecta material. Each packet has properties assigned containing information about the energy and frequencies as well as the direction of the propagation. The initial energy is determined by the relevant radioactive decay processes. The total energy $E_{\rm tot}\left(t_j\right)$ is then divided equally among all the photon packets generated. For the radiation transfer simulations performed, $N_{\rm ph} = 10^6$ photon packets are used.\par 

The initial frequency for each photon packet is determined by Kirchhoff's law, i.e., by sampling over the thermal emissivity:

\vspace{-0.2cm}
\begin{equation}
    S\left(\nu\right) = \kappa\left(\nu\right) B\left(\nu , T \right).
\end{equation}

\noindent Here $\kappa \left(\nu\right)$ is the opacity and $B\left(\nu , T \right)$ is the Planck function at temperature $T$. Thus, the wavelength of the photons depends on the ejecta temperature $T$ and the opacity $\kappa$ of the material, which again depends on the electron fraction $Y_e$. We use the opacities of \cite{Tanaka:2020}. \par 

The photon packets are propagated throughout the ejecta taking into account interactions such as scattering and absorption, that change the properties of the respective photon packet, i.e., the direction, the frequency, and the energy.

Finally, synthetic observable such as flux and polarization spectra are computed using the event-based technique discussed in \cite{Bulla:2019muo} for different observation angles $\theta_{\rm obs}$. For this work, eleven different angles are considered from $\theta_{\rm obs} = 0$ perpendicular to the orbital axis to $\theta_{\rm obs} = \pi / 2$ parallel to the orbital axis in $\Delta \cos \theta_{\rm obs} = 0.1$ steps. We set the azimuth angle $\Phi$ to $0$, i.e., we observe within the $x-z$ plane.

\bibliography{paper20220831.bbl}

\end{document}